# Safety, feasibility, and acceptability of a novel device to monitor ischaemic stroke patients

Samuel J. van Bohemen[1*], Jeffrey M. Rogers[2,3], Aleksandra Alavanja[4], Andrew Evans[4], Noel Young[5], Philip C. Boughton[6,7], Joaquin T. Valderrama[8,9,10,11], Andre Z. Kyme[1,12]

[1] School of Biomedical Engineering, The University of Sydney, Sydney, NSW, Australia
[2] Department of Clinical Medicine, Macquarie University, Sydney, NSW, Australia
[3] Neurocare Group, Sydney, NSW, Australia
[4] Department of Aged Care of Stroke, Westmead Hospital, Sydney, NSW, Australia
[5] Imaging, Western Sydney University, Sydney, NSW, Australia
[6] Sydney Pharmacy School, Faculty of Medicine and Health, The University of Sydney, Sydney, NSW, Australia
[7] Sydney Spine Institute, Sydney, NSW, Australia
[8] Department of Signal Theory, Telematics and Communications, University of Granada, Granada, Spain
[9] Research Centre for Information and Communications Technologies (CITIC-UGR), University of Granada, Granada, Spain
[10] Department of Linguistics, Macquarie University, Sydney, Australia
[11] National Acoustic Laboratories, Sydney, Australia
[12] Brain and Mind Centre, The University of Sydney, Sydney, NSW, Australia

*Corresponding author email address: svan2111@uni.sydney.edu.au
*Corresponding author ORCID: 0000-0001-9620-4099

# Safety, feasibility, and acceptability of a novel device to monitor ischaemic stroke patients

**Abstract**

This study assessed the safety, feasibility, and acceptability of a novel device to monitor ischaemic stroke patients. The device captured electroencephalography (EEG) and electrocardiography (ECG) data to compute an ECG-based metric termed the Electrocardiography Brain Perfusion index (EBPi), which may function as a proxy for cerebral blood flow (CBF). Seventeen ischaemic stroke patients wore the device for nine hours and reported feedback at 1, 3, 6 and 9 hours regarding user experience, comfort, and satisfaction (acceptability). Safety was assessed as the number of adverse events reported. Feasibility was assessed as the percentage of uninterrupted EEG/ECG data recorded (data capture efficiency). No adverse events were reported, only minor incidences of discomfort. Overall device comfort (mean ± 1 standard deviation (SD)) (92.5% ± 10.3%) and data capture efficiency (mean ± 1 SD) (95.8% ± 6.8%) were very high with relatively low variance. The device didn't restrict participants from receiving clinical care and rarely (n=6) restricted participants from undertaking routine tasks. This study provides a promising evidence base for the deployment of the device in a clinical setting. If clinically validated, EBPi may be able to detect CBF changes to monitor early neurological deterioration and treatment outcomes, thus filling an important gap in current monitoring options.

Trial registration: The study was prospectively registered with the Australian New Zealand Clinical Trials Registry (ACTRN12622000112763).

Keywords: electroencephalography; electrocardiography; electrocardiography brain perfusion index; stroke; cerebral blood flow.

## 1. Introduction

Worldwide stroke is the second leading cause of death, the third leading cause of disability and affects approximately 12 million people every year [1]. Although there has been a decrease in global age-standardized incidence rates, the burden of stroke is expected to remain high due to aging populations and increases in cardiovascular risk factors such as obesity, hypercholesteremia, sedentary lifestyles and poor diet [1].

The most common neuroimaging method for stroke diagnosis is computed tomography (CT) scans that measure the perfusion status of brain tissue volumetrically. However, to limit radiation exposure they only provide a "snapshot" around the time of diagnosis and usually again 24 hours later.

Following stroke diagnosis via neuroimaging, patients begin receiving treatment (tissue plasminogen activator (t-PA) or endovascular clot retrieval (EVT)) and are at risk of early neurological deterioration (END), clinical worsening or stroke recurrence within the first 72 hours [2, 3]. END has multiple aetiologies including secondary stroke, t-PA induced symptomatic intracerebral haemorrhage (sICH), delayed cerebral ischaemia (DCI), vasospasm, cerebral edema, and seizures [4-6] and is correlated with poor functional outcome, disability and mortality [7]. Up to 40% of patients with acute ischaemic stroke suffer END [8]. Since almost half of END patients deteriorate within the first 24-48 hours [9, 10], early detection and intervention is vital to prevent serious complications [2].

However, in-between CT scans there is currently no routine continuous monitoring of brain activity and cerebral blood flow (CBF), which are key indicators of stroke occurrence, severity, and patient outcome. Instead, following treatment standard care typically involves high-intensity monitoring of vital signs and neurological status (e.g., National Institutes of Health Stroke Scale (NIHSS), Glasgow Coma Scale (GCS)) every 15 minutes for the first 2 hours, every 30 minutes for the next 6 hours, then hourly up to 24 hours [11]. These protocols are resource intensive, often requiring one-to-one nursing [12] and neurological assessments are subjective [13] and can be influenced by examiner background, experience, and personal bias. Consequently, neurological deficits can go undetected, particularly in patients with limited consciousness [14]. In one study these protocols were only able to neurological deterioration in 45% of patients who experienced neurological deterioration [15].

Electroencephalography (EEG) is a safe, non-invasive method for monitoring brain activity [16]. Continuous EEG is often used to monitor epilepsy but is rarely used to monitor stroke [17]. EEG does not directly monitor CBF, however, quantitative EEG (qEEG) metrics including the delta/alpha ratio (DAR) can be used to detect stroke and inform stroke treatment outcome [18, 19].

Transcranial Doppler (TCD) ultrasound, functional near infrared spectroscopy (fNIRS) and rheoencephalography (REG) can all provide safe, non-invasive, and continuous monitoring of CBF. However, none of these methods are used for routine, long-term monitoring of stroke patients due to limitations. For example, TCD can be used to detect vasospasm and DCI in patients with subarachnoid haemorrhagic stroke but cannot measure small vessel narrowing reliably and has a low sensitivity and specificity [20, 21]. fNIRS is impacted by poor spatial resolution, artefacts and lack of standardization and is therefore mainly used as a research tool [22]. REG is rarely used for clinical applications due to the current lack of pathological and physiological correlations [23]. There is, therefore, opportunity for technological innovation in this space to enable reliable long term CBF monitoring in stroke patients.

Long-term, continuous monitoring of neurologically vulnerable patients in acute settings is challenging and faces technical limitations including quick, stable placement and maintenance of electrodes/probes [24, 25] and the presence of artefacts [26]; and logistical limitations including the lack of 24/7 neurophysiological staff to monitor the device and interpret data [27] and the cost effectiveness of using such systems [28].

For a continuous monitoring system to be successfully integrated into clinical workflows, the system should be safe, minimally invasive, lightweight, wireless, portable, easy to deploy and remove, have a long battery life, be well tolerated by patients and not cause electrical interference with other hospital equipment or interfere with routine care [27].

An electrocardiography (ECG)-based metric, termed the Electrocardiography Brain Perfusion index (EBPi), has been developed to monitor changes in CBF [29]. EBPi is based on the hypothesis that changes in CBF will alter the propagation of the ECG signal from the chest to the scalp, hence altering the appearance of this signal at scalp electrodes with respect to the same signal recorded across the chest. A feasibility study in healthy volunteers demonstrated that EBPi correlated with TCD and may function as a proxy for CBF changes [29].

The aim of this study is to assess the safety, feasibility, and acceptability of a new device to monitor EBPi in ischaemic stroke patients during recovery in hospital. We hypothesized the device would be safe, capable of capturing reliable data and well tolerated by patients.

## 2. Methods

### 2.1. Study design

A prospective, single site observational study was undertaken to assess the safety, feasibility, and acceptability of a novel device to monitor ischaemic stroke patients. The site, Westmead Hospital, Sydney, is a Level 6 adult University Referral Hospital with a 24-hour acute stroke service that receives over 400 ischaemic stroke patients annually.

The study design was based on safety, feasibility, and acceptability assessments of similar brain monitoring devices [30-35]. All experimental procedures were conducted in accordance with an approved human ethics protocol (WSLHD 2021/ETH11420) and were performed in accordance with the ethical standards stated in the 1964 Declaration of Helsinki and its later amendments. The study was prospectively registered with the Australian New Zealand Clinical Trials Registry (ACTRN12622000112763). Primary outcomes are listed below:

- Safety: Assessed via reported adverse events.
- Feasibility: Assessed for each scalp and chest electrode as the percentage of uninterrupted EEG/ECG data recorded (data capture efficiency).
- Acceptability: Assessed via questions focusing on user experience, comfort, and satisfaction.

Details of each aspect are described in the following sections.

### 2.2. Device construction

Three lightweight, portable devices (Figure. 1a) were constructed to wirelessly record EBPi data in ischaemic stroke patients. Each device comprised a 3D-printed thermoplastic polyurethane band housing four snap ECG electrodes connected to four neonatal ECG sticker electrodes (disposable electrode PG10C, FIAB) (Figure. 1b), two snap ECG electrodes connected to adult ECG sticker electrodes (WhiteSensor WS/RT, Ambu), two ground/reference ear clip electrodes (Figure. 1c), an elastic fabric headband with a Velcro strap and an 8-channel 250 Hz sampling OpenBCI Cyton Board (Cyton Biosensing Board, OpenBCI) to capture bioelectrical signals (Figure. 1d). A 3D printed polylactic acid enclosure (electronics housing) (Figure. 1e) housed the OpenBCI Cyton Board, a small lithium-ion battery (3.7 V, 500 mA) and a charging unit.

## *2.3. Participants*

Seventeen patients (mean age ± 1 standard deviation (SD) = 67.4 yr ± 13.1 yr, range 33–86 yr) admitted to the Westmead Hospital stroke ward between March-July 2022 with clinical and radiologically confirmed ischaemic stroke (either from CT or magnetic resonance imaging (MRI)) participated in the study. Sample size was based on previous clinical safety studies [30-35].

Ischaemic stroke was classified using the Oxfordshire classification system (total anterior circulation infarcts (TACI), partial anterior circulation infarcts (PACI), lacunar circulation infarcts (LACI), and posterior circulation infarcts (POCI)) [36]. Participants had received treatment (if appropriate) and were to be in the stroke ward for recovery and monitoring for at least 9 hours. The clinical team was responsible for selecting eligible participants and coordinating informed consent. Exclusion criteria included: haemorrhagic stroke, stroke mimic, transient ischaemic attack, or no stroke; brain surgery or any intervention or condition that would prevent the participant from wearing the device for 9 hours; having already been monitored using EEG equipment; having a history of epilepsy and/or seizures; brain or scalp implants; cognitive impairment making a participant more susceptible to forms of discomfort or distress caused by wearable devices; unable to complete the questionnaire by themselves or with the help of a family member, guardian, caregiver or nurse. Eligible participants unable to provide consent were included in the study if an authorized guardian or carer could provide consent. Participant recruitment was based on eligibility and was not consecutive.

*2.4. Device deployment*

A nurse applied No-Sting-Barrier-Film (Cavilon™, 3M, St Paul, MN) to the forehead and upper chest of participants to prevent skin abrasion from the ECG sticker electrodes. The head-worn device was then fitted such that the yellow midline (Figure. 1c) was centred on the forehead, the black band was situated above the eyebrows and the neonatal ECG sticker electrodes contacted the forehead (Fp1, Fp2, F7 and F8) (Figure. 2). The elastic fabric headband was adjusted to ensure a comfortable yet firm fit to the head using the Velcro strap. The adult ECG sticker electrodes were placed below each clavicle at LA and RA and the ground/reference electrodes placed on the ear lobes (Figure. 2). Synchronized EEG and ECG data were streamed wirelessly from the device to a nearby laptop via the OpenBCI RF USB dongle for the duration of the study. The data stream was monitored in real-time from a nearby room.

*2.5. Questionnaire*

A questionnaire (Figure. 3) was deployed at 1, 3, 6 and 9 hours to evaluate device safety and acceptability. Participant answers were ranked using a 9-point scale (1=extremely negative to 9=extremely positive). Other questions required a yes/no answer with an explanation, including four additional questions at the conclusion of the 9-hour session where participants were asked if they would be willing to wear a similar device in hospitals and at home to monitor their brain health and detect future stroke. Participants were also asked if they used any other wearable medical devices.

*2.6. Clinical care*

During the 9-hour recording session the participants received normal clinical care, including monitoring of blood pressure, ECG and oxygen saturation, routine neurological assessments (GCS), clinician consults, allied health consults (physiotherapy, speech pathology) and any other routine tests (e.g., follow-up CT, MRI scans). If participants were to have a physiotherapy session requiring them to walk around, the physiotherapist pushed the recording computer on a trolley alongside to ensure it remained within 2–3 meters of the participant. The device was temporarily removed if participants required any imaging or other procedure requiring them to leave the stroke ward; in this case the study clock was paused during the procedure and then recording resumed once the device was re-fitted to the participant back in the ward. Within the ward, participants were free to move around, visit the bathroom, eat, and sleep without the device being removed. Instances of participant movement that were observed by the research team were documented. However, since participants were not observed continuously, undocumented participant movement was likely.

*2.7. Data analysis*

*2.7.1. Safety*

Device safety was quantified based on the cumulative number of adverse events and near adverse events reported during a 9-hour session. As defined by the Therapeutics Goods Administration, Australia, an adverse event is an occurrence involving a medical device that meets the following criteria:

- Death of a patient, health care provider, user, or other person; or
- A serious injury or serious deterioration to a patient, health care provider, user, or other person, including:
    - A life-threatening illness or injury;
    - Permanent impairment of a body function;
    - Permanent damage to a body structure; or
    - A condition necessitating medical or surgical intervention to prevent permanent impairment of a body function or permanent damage to a body structure.

A near adverse event is an occurrence involving a medical device that might have led to death or serious injury [37].

All incidences of discomfort from wearing the device were reported in answers to Q9.

*2.7.2. Feasibility*

Answers to Q7 reported when and what electrodes came loose to identify interrupted EEG/ECG data which was later confirmed by visual inspection of the data. For each participant, the number of uninterrupted EEG/ECG data samples were then calculated as percentage of the total number of recorded samples to compute data capture efficiency at each scalp (Fp1, Fp2, F7 and F8) and chest (LA and RA) electrode. For each electrode, mean data capture efficiency was computed by averaging across all participants. An overall data capture efficiency was computed by averaging across all participants and electrodes.

For each participant, EBPi data were computed using the recorded EEG/ECG data as outlined in [29].

*2.7.3. Acceptability*

For each participant, a device comfort score was computed by averaging the numeric responses to Q1–6 and Q10 at 1, 3, 6 and 9 hours. A mean device comfort score at the different time intervals was computed by averaging these values across all participants. An overall device comfort was computed by averaging scores across all participants and time intervals.

To assess if the device restricted normal activities, numeric responses to Q8 were averaged across participants at the different time intervals and the restricted activities were recorded. An overall score was also computed by averaging across all participants and time intervals.

**3. Results**

*3.1. Participant data*

Participant demographics, neuroimaging, stroke diagnosis, treatment regime, clinical imaging and other activities during the study are shown in Table 1 for *n*=17. The cohort represented 8 different ethnicities who presented mainly with PACI and POCI (82%). More than half of the participants received no stroke treatment post-diagnosis.

Of the 17 participants, 12 were able to answer the questionnaire on their own, 4 required the aid of a family member and 1 participant required a nurse/hospital translator to convey the questions due to a language barrier.

*3.2. Safety*

No adverse events or near adverse events were reported for any participant. However, there were seventeen incidences of discomfort reported, including scalp discomfort ($n=3$), minor headache ($n=4$), ear discomfort caused by the electronics housing rubbing on the right ear (housing-ear impingement) ($n=4$), ear lobe discomfort caused by the reference ear clip electrodes ($n=4$), and hair caught in the Velcro strap ($n=2$) (Figure. 4a). The total number of incidences of discomfort reported at 1, 3, 6 and 9 hours is shown in Figure. 4b. Over half of the incidences of discomfort were reported at 9 hours. The degree of discomfort (mean ± 1 SD) caused by these incidences was quantified using the 9-point scale and is displayed in Figure. 5. The equal highest degree of discomfort was caused by scalp discomfort, minor headache, and housing-ear impingement. There were no significant differences in the degree of discomfort caused by any of the reported incidences (two-tailed Student's t-test, $p<0.05$).

*3.3. Feasibility*

Data capture efficiency (mean ± 1 SD) for the scalp and chest electrodes are displayed in Figure. 6. Scalp electrode F7 had the lowest data capture efficiency and the highest variance 90.4% ± 13.0% (range 54.8%-100%). A significant difference (two-tailed Student's t-test, $p<0.05$) was detected between data capture efficiency at scalp electrode F7 and scalp electrode Fp2, chest electrodes LA and RA. Mean data capture efficiency (mean ± 1 SD) (across all electrodes) for each participant is shown in Figure. 7. Participant 008 had the lowest data capture efficiency 88.7% ± 16.6%. Overall data capture efficiency (across all electrodes and participants) was 95.8% ± 6.8% (range 54.8%–100%).

EBPi data from one participant is shown in Figure. 8 to illustrate how future studies might explore the sensitivity of this metric to detect and monitor ischaemic stroke patients. There were several large changes in EBPi during the 9-hour session, many of which coincided with either the documented participant movement (e.g., physiotherapy session at 80 mins) or obvious changes in accelerometery data captured from the device (e.g., at 230 mins).

*3.4. Acceptability*

Device comfort scores (mean ± 1 SD) at 1, 3, 6 and 9 hours are shown in Figure. 9. The mean device comfort score never dropped below 8.3/9 (92.0%) at a given time interval. No significant differences were detected across the time internals (two-tailed Student's t-test, $p<0.05$). The overall device comfort score (across all time intervals) was 8.3/9 (92.5% ± 10.3%, range 5.1/9–9/9 (57.1%–100%)).

Mean numeric answers to Q8 (mean ± 1 SD) at 1, 3, 6 and 9 hours are shown in Figure. 10. This score was ≥ 8.4/9 (93.5%) at all time intervals, and variance was typically ~10%. No significant differences were detected across the time internals (two-tailed Student's t-test, $p<0.05$). The overall score (across all time intervals) was 8.7/9 (96.4% ± 9.9%, range 5/9-9/9 (55.6%-100%)). The device was reported to interfere with routine tasks 6 times. The study was paused 11 times in total (Table 1, clinical/other activities during study).

*3.5. Additional questions*

All participants (n=17) said they would be willing to wear a similar device in hospital and at-home to monitor their brain health and to detect future stroke. In response to "Do you normally use any other wearable medical devices?", 2 participants reported wearing hearing aids and 1 participant reported wearing a smart watch.

## 4. Discussion

Numerous studies have investigated the feasibility and acceptability of head-worn devices for long-term, continuous monitoring of neurological disorders including epilepsy [27, 30, 38]. However, there is little reported on the utility of such devices for long-term, continuous monitoring of stroke. Therefore, this study introduces a new device to monitor ischaemic stroke patients during hospital recovery and investigated the safety, feasibility, and accessibility of the device. The lightweight, portable device captures EEG and ECG signals to compute a novel metric associated with changes in CBF [29].

*4.1. Safety*

No adverse events or near adverse events were reported from a total of 153 hours of device usage in the stroke ward. However, seventeen incidences of discomfort were reported (Figure. 4). According to the 2018 NHRMC National Standard 2.1.6: research is "low risk" where the only foreseeable risk is one of discomfort [39]. Therefore, this result can be considered an acceptable risk. The mean degree of discomfort caused by these incidences was never below 5/9 (55.6%) (Figure. 5), revealing the discomfort level was tolerable but confirms the design can be further improved for comfort.

When these incidences of discomfort were reported, steps were taken to ameliorate them. Scalp discomfort and minor headache may have been caused by the device being too tight, although it is possible that ischaemic stroke may have also contributed towards the presentation of headache [40]. Loosening the Velcro strap relieved these issues. Repositioning the electronics housing higher on the elastic fabric headband (away from the right ear) helped prevent housing-ear impingement. When ear lobe discomfort was reported, the ground/reference ear clips were repositioned to a new location on the same ear. Certain hair types were prone to getting caught in the Velcro strap. Careful device placement reduced this risk.

*4.2. Feasibility*

Overall data capture efficiency was very high (95.8% ± 6.8%), indicating the device successfully captured EEG/ECG data during most of the study. Data loss was caused by multiple factors including poor scalp/chest/ear electrode contact, poor connection between the electrode leads and OpenBCI Cyton Board (Figure. 1e, black plug) and an interrupted data stream. If the researcher monitoring the data steam detected data loss at one or more electrodes, the signal was restored as soon as possible without interfering with routine clinical care. At times there was a delay before the signal could be restored, this contributed to reduced data capture efficiency scores. In clinical practice, it is unlikely that someone would monitor the data steam continuously. Instead, an automated monitoring system could be implemented to detect and alert to poor electrode contact and data loss.

Scalp electrodes came loose due to two main reasons: 1) the elastic fabric headband was too loose and 2) overtime the elastic fabric headband rubbed on the participants pillow causing the headband to move up the back of the participants head pulling the scalp electrodes of the participants forehead. This was mitigated by ensuring the Velcro strap was tight enough to prevent the device from moving. If the device had moved off the participants head, the device was repositioned ensuring a tight fit.

Scalp electrodes F7 had the lowest data capture efficiency (90.4%) and highest variance (± 13.0%) and scalp electrode F8 had the second lowest data capture efficiency (94.6%) and second highest variance (± 6.3%) (Figure. 6). This is likely because F7 and F8 are the most peripheral scalp electrodes (Figure. 2), hence were the first to lose contact with the scalp if the elastic fabric headband was too loose and/or the device started to come off the participant's head.

The chest electrodes had the highest data capture efficiency (both 98.0%) and lowest variance (both ± 2.1%) (Figure. 6). Chest electrodes came loose due to hair on the participants chest. When this occurred, the electrode(s) was repositioned, when possible, to a less hairy region.

The ground/reference ear clip electrodes came loose due to a clip electrode slipping off the participants ear lobe, hitting the pillow, while the participant was on the phone and one-time a nurse accidentally removed it. When this occurred, the clip electrode was reattached as soon as possible.

When the connection between the electrode leads and the OpenBCI Cyton Board (Figure. 1e, black plug) came loose, the signal at one or more electrodes was lost. When this occurred, the black plug was reconnected ensuring a secure connection.

A poor Bluetooth connection also caused the data stream to be interrupted, causing data loss. A few times the data stream was interrupted due to a reason that is yet to be confirmed. When either of these events occurred the data stream was restarted.

Four participants pulled the device off completely. One participant removed the device because they wanted to go to the bathroom; one participant was disorientated; one participant wanted to sleep on their right-hand side and said the electronics housing prevented them from doing so; one participant was agitated. If a participant removed the device, the research team re-deployed the device as soon as possible. There were no instances where the research team were unable to successfully re-deploy the device.

On average it took less than 5 mins to deploy and remove the device. In some participants one or more neonatal ECG sticker electrode(s) remained on the forehead after device removal but they were subsequently removed with ease.

*4.3. Acceptability*

The overall device comfort score was very high (92.5% ± 10.3%), indicating that the device was very well tolerated. The participant that pulled the device off, reporting the device prevented them from sleeping on their right-hand side, gave the lowest comfort score of 57.1%. The device did not restrict participants from receiving normal clinical care and did not cause electrical interference with hospital equipment.

The overall score for Q8 (To what extent did wearing the device restrict your normal everyday roles and activities? (1= extremely restrictive, 9=not restrictive at all)) was also very high (96.4% ± 9.9%), indicating that the device rarely (*n*=6) restricted participants from doing normal everyday roles and activities. Four participants reported that the electronics housing prevented them from sleeping on their right-hand side. One participant reported the device made it tricky for them to get dressed. One nurse reported the device interfered with the process of rolling a participant over to be cleaned.

*4.4. Limitations*

Although the study provides strong evidence that the device was safe, able to capture real data, and was very well tolerated by participants, there were a number of limitations. Questionnaires are subjective and prone to missing information. The clinical team determined that all participants had the capacity to receive information and respond to the questionnaire with or without the help of a family member, guardian, caregiver, or nurse. However, because participants had recently had an ischaemic stroke, it is possible that they were not in a stable condition to provide real, accurate answers. Furthermore, the questionnaire was created for the study and is not a validated questionnaire.

There is also a possibility of confirmation bias, where a respondent (participant) is inclined to agree with existing beliefs or expectations, in this case, that the device is safe and will be well tolerated by participants [41]; and authority bias, where a respondent is influenced by the opinion of an authority figure (i.e., a nurse or clinician) [42].

Because of the small sample size ($n$=17) and exclusion criteria, the participant cohort is unlikely to be a true representation of ischaemic stroke. Different findings may be expected with a larger cohort that truly represents ischaemic stroke.

The participants suffered an ischaemic stroke several days prior to the study (mean days from stroke onset to study was 4.2, SD 2.9, range 1–9). It is possible participant responses may have been different if the device was deployed immediately following stroke diagnosis. This will be explored in a future study.

During the present study, EBPi was not expected to detect the presence of stroke, as the stroke events occurred several days earlier, and participants had already received treatment (if appropriate) to restore CBF. Five participants received t-PA, two participants received EVT, and ten participants received no treatment.

EBPi data displayed large changes that could be related to participant activity and movement. Eating and sleeping are known to induce changes in CBF [43, 44]. It is likely that these activities went undocumented during the study and may have impacted the EBPi data. The OpenBCI Cyton Board contains a 3-axis accelerometer. Analysis of this accelerometry data overlaid with the EBPi data shows changes in EBPi at the same time as accelerometery detected participant movement (Figure. 8). Head movement and changes in posture or body orientation (i.e., supine, standing) are also known to induce changes in CBF [45, 46], which could also have impacted the EBPi data. Future studies will explore the relationship between eating, sleeping, head/participant movement and EBPi to develop an algorithm to produce reliable EBPi data during participant activity and movement.

*4.5. Future direction*

Clinical safety studies provide important safety and useability data to inform device development and further clinical trials before a device is clinically available. Findings from this study will be used as important inputs for the next device iteration: 1) A new elastic fabric headband design will be considered to prevent device movement caused by rubbing on the user's pillow, 2) the electronics housing will be redesigned and repositioned to prevent ear impingement and to allow users to sleep on their side, 3) the connection between the electrodes leads and the OpenBCI Cyton Board could be made permanent, 4) users' chest hair could be shaven to ensure stable chest electrode contact, 5) repositioning of the ground/reference electrode will be considered to mitigate lost contact and 6) the Velcro strap could be replaced for another fastening mechanism.

This study demonstrated that the device could monitor EEG, ECG and EBPi data continuously over 9 hours in ischaemic stroke patients, who currently have no real-time, quantitative monitoring of brain activity or CBF. A clinical trial is planned to monitor patients with large vessel occlusion, anterior circulation ischaemic stroke. The device will be deployed as soon as possible following radiologically confirmed ischaemic stroke and will be worn for 24 hours. The aim of this trial is to determine if EBPi can detect 1) the presence and location of stroke, 2) treatment outcome and 3) END. Future trials will also use EBPi in combination with stroke sensitive qEEG metrics like DAR [18] to provide simultaneous monitoring of CBF and brain activity; to increase the sensitivity to stroke, and to develop an identification pattern that can be used to detect stroke, monitor treatment and track END.

The initial 24 hours after t-PA administration are critical. t-PA induced sICH occurs in 6% of patients and carries a 50% chance of death [6]. Since END is defined as clinical worsening or stroke recurrence within the first 72 hours [2, 3], continuous monitoring of patients during this time may add the most value to stroke patient management. However, because nursing observations are less frequent during days 4–7, continuous monitoring of patients during this time may also be beneficial. A device that can provide real-time monitoring of patients following t-PA administration to monitor treatment outcome and detect END such as sICH in real-time, would enable patients to receive intervention faster than current methods allowing for improved patient outcomes. It could also reduce the burden on nursing staff to perform neurological assessments.

## 5. Conclusion

This study assessed the safety, feasibility, and acceptability of a novel device to monitor a small cohort of ischaemic stroke patients. No adverse events were reported, the device reliably captured assessable EEG, ECG and EBPi data, the device was well tolerated by participants and was able to be (re)deployed with ease. Findings from patient feedback on comfort will provide key input for future iterations of the device. A follow up clinical trial is planned to assess the clinical utility of EBPi for the continuous monitoring of ischaemic stroke patients. If clinically validated, EBPi may be able to provide continuous monitoring of CBF to monitor treatment outcomes and detect END.

## Declaration of interest statements

### *Disclosure of interest*

SJVB is a co-founder and shareholder of Nuroflux Pty Ltd. PCB is a shareholder of Nuroflux Pty Ltd. Nuroflux Pty Ltd is the applicant of a published PCT application (WO2021/077154A1) for the described method of monitoring changes in CBF. SJVB, PCB and AZK are listed as inventors on the published PCT application.

### *Data availability statement*

The participants of this study did not give written consent for their data to be shared publicly, so due to the sensitive nature of the research supporting data is not available.

# Tables

**Table 1.** Participant demographics, stroke presentation, clinical/other activities and hospital stay.

| | |
|---|---|
| Age [a] | 67.4 (13.1), 33–86 |
| Gender | |
|   Male | 10 (59%) |
|   Female | 7 (41%) |
| Ethnicity | |
|   Australian | 8 (47%) |
|   Aboriginal or Torres Strait Islander | 1 (6%) |
|   Chinese | 2 (12%) |
|   Indian | 2 (12%) |
|   Ukraine | 1 (6%) |
|   Lebanese | 1 (6%) |
|   Tongan | 1 (6%) |
|   Greek | 1 (6%) |
| Neuroimaging modality to confirm stroke | |
|   NCCT, CTP, CTA | 10 (59%) |
|   NCCT | 2 (12%) |
|   MRI | 5 (29%) |
| Stroke presentation | |
|   TACI | 1 (6%) |
|   PACI | 9 (53%) |
|   LACI | 2 (12%) |
|   POCI | 5 (29%) |
| NIHSS on admission [a] | 6.1 (6.8), 1–25 |
| Treatment | |
|   t-PA | 5 (29%) |
|   EVT | 2 (12%) |
|   No treatment | 10 (59%) |
| Clinical imaging during study | |
|   CT scan | 3 |
|   MRI | 2 |
|   X-ray | 1 |
|   Carotid ultrasound | 1 |
|   Echocardiography | 2 |
| Other activities during study | |
|   Shower | 1 |
|   Prayers | 1 |
| Days from stroke onset to study [a] | 4.2 (2.0), 1–9 |
| Days from study to discharge [a] | 9.1 (13.7), 1–52 |
| Total days in hospital [a] | 13.3 (14.6), 3–59 |
| Discharged | |
|   Home | 16 (94%) |
|   Rehab | 1 (6%) |

[a] Mean (standard deviation), range. Abbreviations: CT: Computed tomography; NCCT: Non-contrast computed tomography; CTP: Computed tomography perfusion; CTA: Computed tomography angiography; MRI: Magnetic resonance imaging; TACI: Total anterior circulation infarcts; PACI: Partial anterior circulation infarcts; LACI: Lacunar circulation infarcts; POCI: Posterior circulation infarcts; t-PA: Tissue plasminogen activator; EVT: Endovascular clot retrieval.

**Figures**

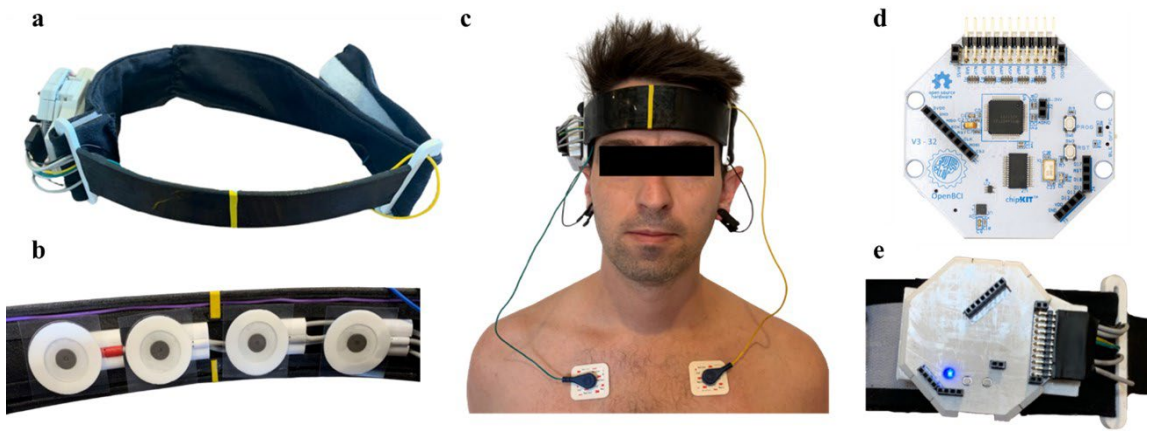

**Figure. 1.** Device to capture EBPi (a) across four neonatal ECG sticker electrodes (b) placed across the scalp, with two adult ECG sticker electrodes placed across the chest at LA and RA (c). OpenBCI Cyton Board (d) housed inside a 3D printed enclosure (electronics housing) secured to the side of the device (e).

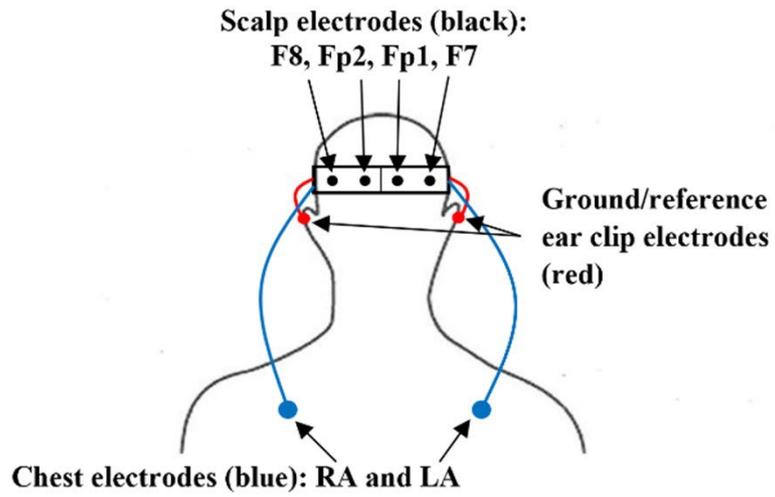

**Figure. 2.** Device and electrode placement across the scalp (Fp1, Fp2, F7 and F8), chest (RA and LA) and ears.

| Use the following 9-point scale to answer the questions below:  1 = Extremely negative  2 = Very negative  3 = Negative  4 = Slightly negative  5 = Neutral  6 = Slightly positive  7 = Positive  8 = Very positive  9 = Extremely positive ||||||
|---|---|---|---|---|---|
| **Unique Participant Study Number:** | | | | | |
| **Questions (The following 10 questions will be answered after 1, 3, 6 and 9 hours)** | | **1 hour** | **3 hours** | **6 hours** | **9 hours** |
| 1. | How comfortable are the scalp electrodes? (1 = extremely uncomfortable, 9 = extremely comfortable) | | | | |
| 2. | How comfortable are the chest electrodes/wires? (1 = extremely uncomfortable, 9 = extremely comfortable) | | | | |
| 3. | How comfortable are the ear clips/wires? (1 = extremely uncomfortable, 9 = extremely comfortable) | | | | |
| 4. | How comfortable is the headband? (1 = extremely uncomfortable, 9 = extremely comfortable) | | | | |
| 5. | How comfortable is the temperature of the device (1 = extremely uncomfortable, 9 = extremely comfortable) | | | | |
| 6. | How comfortable do you feel wearing the device? (1 = extremely uncomfortable, 9 = extremely comfortable) | | | | |
| 7. | Did any of the electrodes come loose? (Yes/No) If yes, what electrodes? | | | | |
| 8. | To what extent did wearing the device restrict your normal everyday roles and activities? (1= extremely restrictive, 9=not restrictive at all) Which roles and activities? | | | | |
| 9. | Did you experience any adverse effects from the device (i.e., Scalp discomfort, skin iteration, itching, tingling, headaches etc.)? (Yes/No) If yes, what? | | | | |
| 10. | Overall, how satisfied were you with the device? (1 = extremely unsatisfied, 9 = extremely satisfied) | | | | |
| **Questions (The following 4 questions will be answered at end of the study)** | | **End of the study** | | | |
| 1. | In the future, would you be happy to wear a device like this in hospital to monitor your brain health? (Yes/No) Any comments? | | | | |
| 2. | Would you wear a device like this at home to monitor your brain health? (Yes/No) Any comments? | | | | |
| 3. | Do you normally use any other wearable medical devices? (Yes/No) If yes, what is your satisfaction with this device? | | | | |
| 4. | Do you have any other comments or suggestions about our device? | | | | |

**Figure. 3.** Questionnaire deployed at 1, 3, 6 and 9 hours.

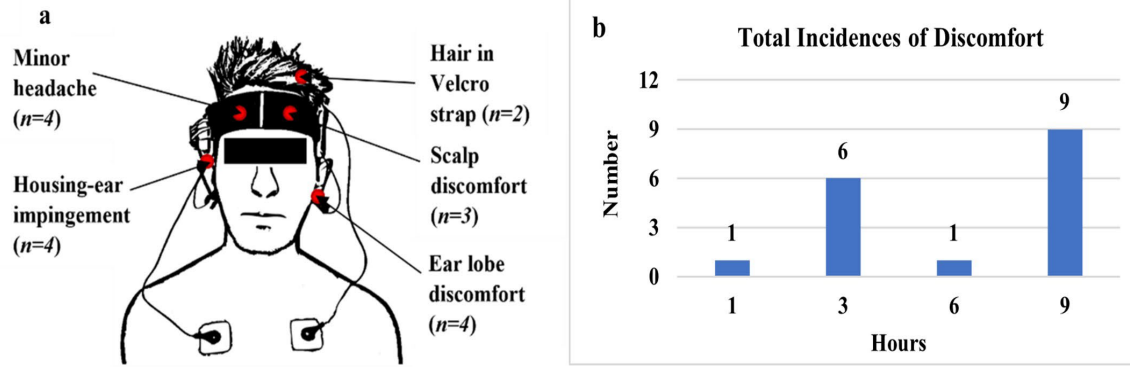

**Figure. 4.** Reported incidences of discomfort (a) at 1, 3, 6 and 9 hours (b).

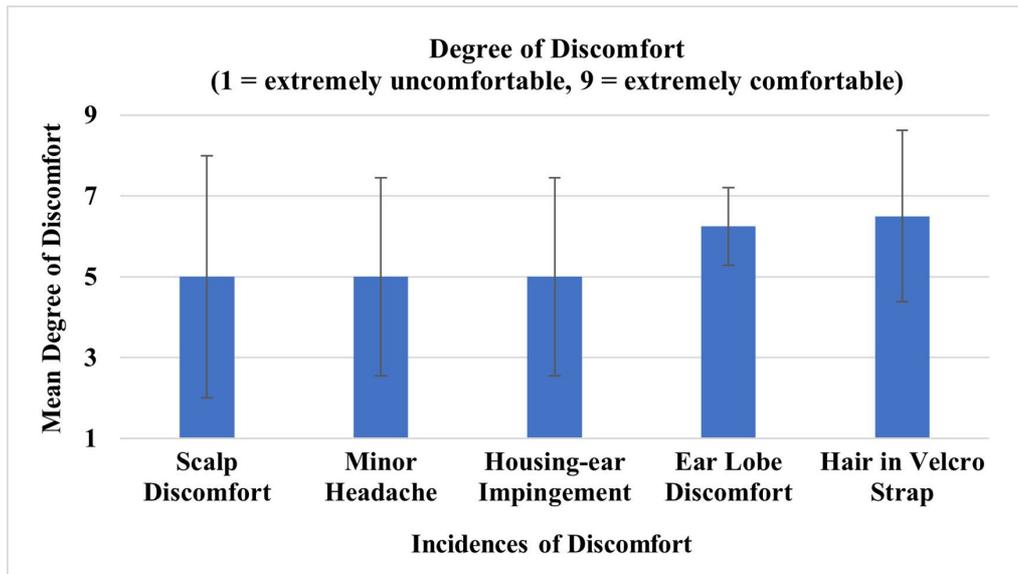

**Figure. 5.** Mean degree of discomfort caused by minor incidences. Error bars show ± 1 standard deviation.

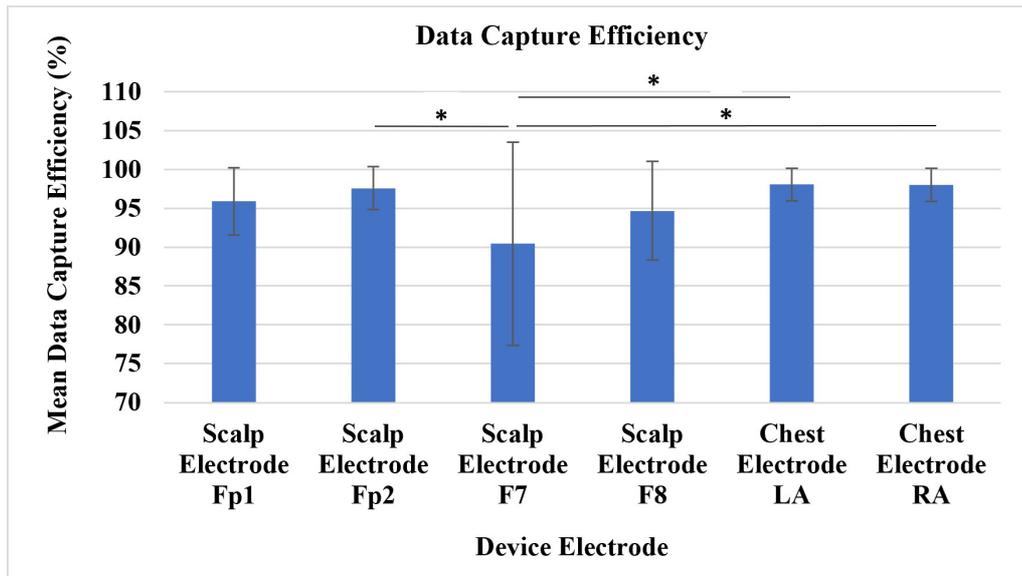

**Figure. 6.** Mean data capture efficiency for each scalp (Fp1, Fp2, F7 and F8) and chest (RA and LA) electrode. Error bars show ± 1 standard deviation. **\*** = Significant difference p<0.05.

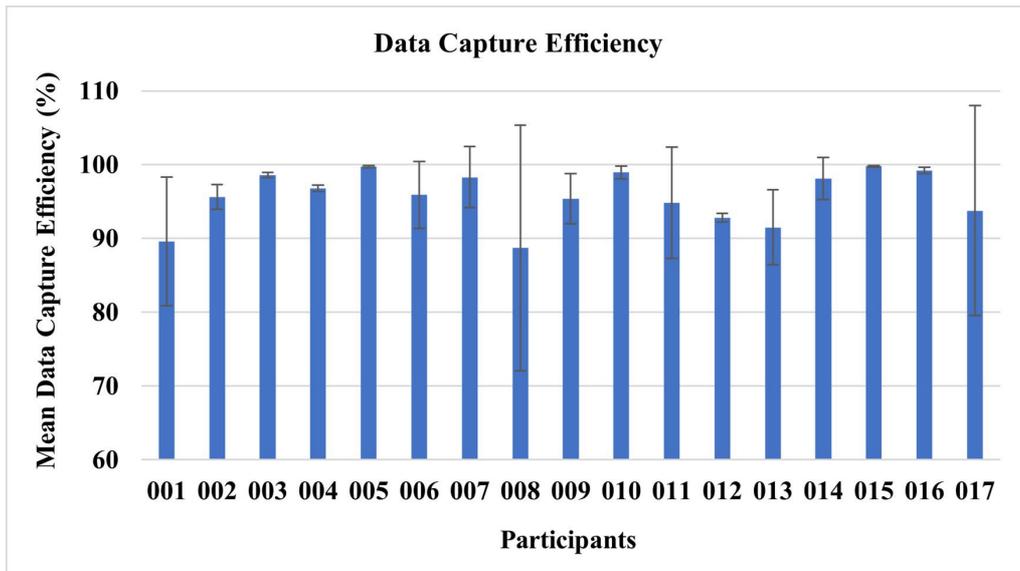

**Figure. 7.** Mean data capture efficiency (averaged across all electrodes) for each participant. Error bars show ± 1 standard deviation.

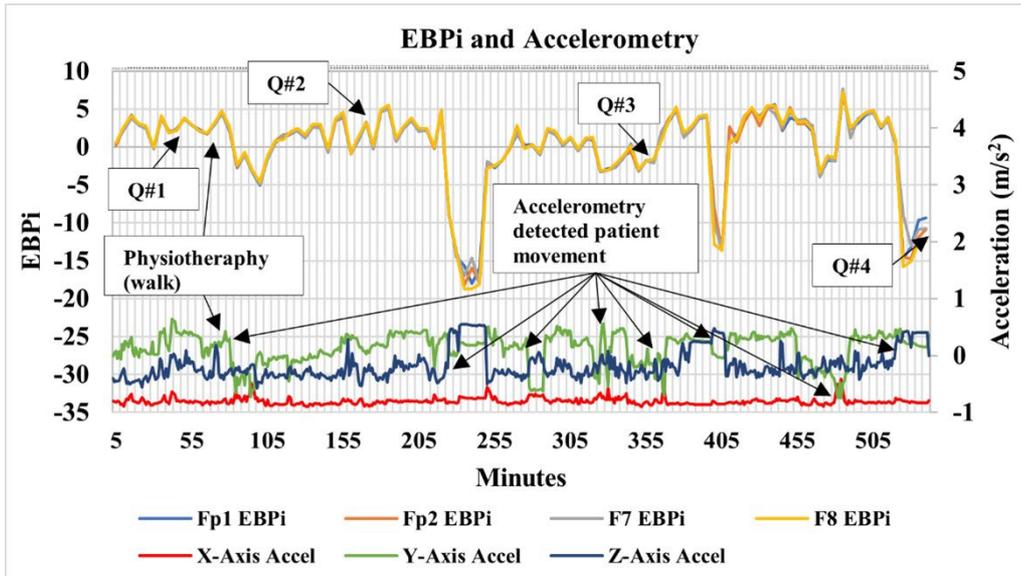

**Figure. 8.** EBPi and device accelerometery data captured from one participant. EBPi data captured at scalp electrodes Fp1, Fp2, F7 and F8 is displayed on the top portion of the figure, 3-axis accelerometery data is displayed on the bottom portion of the figure. Q#: Questionnaire deployed; Accel: Acceleration. Note: EBPi data shows very similar changes all at scalp electrodes therefore it is hard to distinguish the individual EBPi traces at the four scalp electrodes.

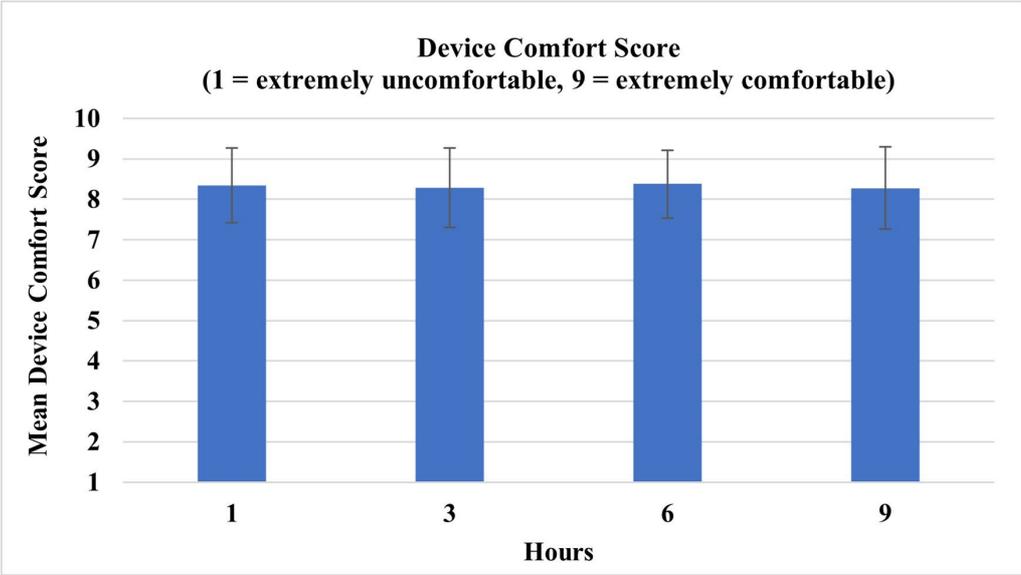

**Figure. 9.** Mean device comfort score at 1, 3, 6 and 9 hours. Error bars show ± 1 standard deviation.

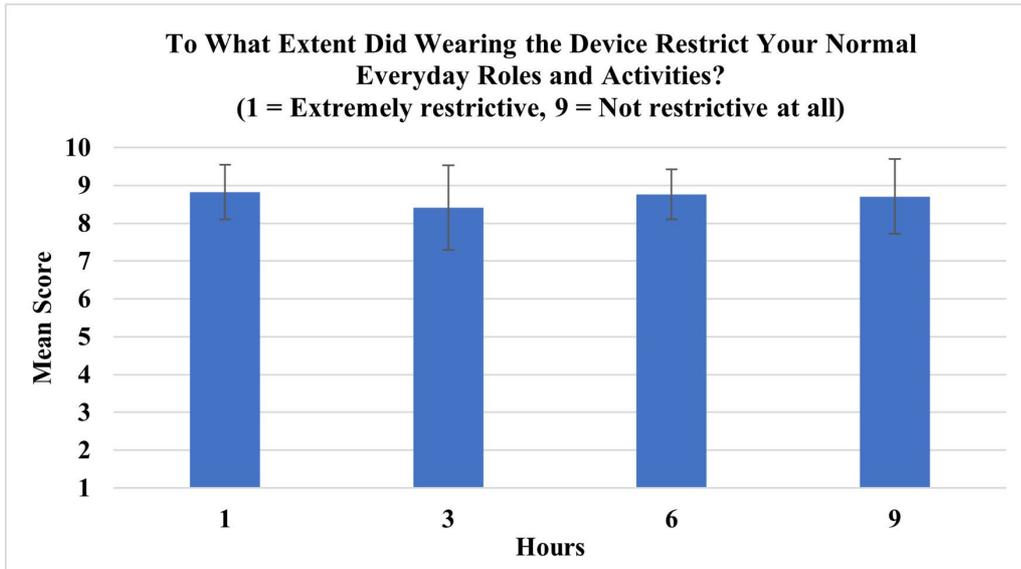

**Figure. 10.** Mean response scores to Q8: To what extent did wearing the device restrict your normal everyday roles and activities? (1= extremely restrictive, 9=not restrictive at all) at 1, 3, 6 and 9 hours. Error bars show ± 1 standard deviation.

**Figure captions**

**Figure. 1.** Device to capture EBPi (a) across four neonatal ECG sticker electrodes (b) placed across the scalp, with two adult ECG sticker electrodes placed across the chest at LA and RA (c). OpenBCI Cyton Board (d) housed inside a 3D printed enclosure (electronics housing) secured to the side of the device (e).

**Figure. 2.** Device and electrode placement across the scalp (Fp1, Fp2, F7 and F8), chest (RA and LA) and ears.

**Figure. 3.** Questionnaire deployed at 1, 3, 6 and 9 hours.

**Figure. 4.** Reported incidences of discomfort (a) at 1, 3, 6 and 9 hours (b).

**Figure. 5.** Mean degree of discomfort caused by minor incidences. Error bars show ± 1 standard deviation.

**Figure. 6.** Mean data capture efficiency for each scalp (Fp1, Fp2, F7 and F8) and chest (RA and LA) electrode. Error bars show ± 1 standard deviation. * = Significant difference $p<0.05$.

**Figure. 7.** Mean data capture efficiency (averaged across all electrodes) for each participant. Error bars show ± 1 standard deviation.

**Figure. 8.** EBPi and device accelerometery data captured from one participant. EBPi data captured at scalp electrodes Fp1, Fp2, F7 and F8 is displayed on the top portion of the figure, 3-axis accelerometery data is displayed on the bottom portion of the figure. Q#: Questionnaire deployed; Accel: Acceleration. Note: EBPi data shows very similar changes all at scalp electrodes therefore it is hard to distinguish the individual EBPi traces at the four scalp electrodes.

**Figure. 9.** Mean device comfort score at 1, 3, 6 and 9 hours. Error bars show ± 1 standard deviation.

**Figure. 10.** Mean response scores to Q8: To what extent did wearing the device restrict your normal everyday roles and activities? (1= extremely restrictive, 9=not restrictive at all) at 1, 3, 6 and 9 hours. Error bars show ± 1 standard deviation.